\documentclass{article}

\usepackage{PRIMEarxiv}
\usepackage{booktabs} 
\usepackage{array}
\usepackage{multirow}
\usepackage{enumitem}
\usepackage{colortbl}
\usepackage{wasysym}
\usepackage{xcolor}
\usepackage[normalem]{ulem}

\usepackage[utf8]{inputenc} 
\usepackage[T1]{fontenc}    
\usepackage{hyperref}       
\usepackage{url}            
\usepackage{booktabs}       
\usepackage{amsfonts}       
\usepackage{nicefrac}       
\usepackage{microtype}      
\usepackage{lipsum}
\usepackage{fancyhdr}       
\usepackage{graphicx}       
\graphicspath{{media/}}     

\usepackage{booktabs} 
\usepackage{array}
\usepackage{multirow}
\usepackage[export]{adjustbox}
\usepackage{pifont}
\usepackage{subfigure}
\usepackage{enumitem}

\usepackage[normalem]{ulem}
\usepackage{setspace}

\title{Emerging Practices for Large Multimodal Model (LMM) Assistance for People with Visual Impairments: Implications for Design
}

\author{
  Jingyi Xie\textsuperscript{1}, 
  Rui Yu\textsuperscript{2},
  He Zhang\textsuperscript{1}, 
  Sooyeon Lee\textsuperscript{3}, 
  Syed Masum Billah\textsuperscript{1}, 
  John M. Carroll\textsuperscript{1} 
  \\
  {1} College of Information Sciences and Technology, Pennsylvania State University
  \\
  {2} Department of Computer Science and Engineering, J.B. Speed School of Engineering, University of Louisville
  \\
  {3} Department of Informatics, Ying Wu College of Computing, New Jersey Institute of Technology
}

\newcommand{\bma}{Be My AI}
\newcommand{\bme}{Be My Eyes}

\begin{document}
\maketitle

\begin{abstract}
People with visual impairments perceive their environment non-visually and often use AI-powered assistive tools to obtain textual descriptions of visual information.
Recent large vision-language model-based AI-powered tools like Be My AI are more capable of understanding users' inquiries in natural language and describing the scene in audible text; however, the extent to which these tools are useful to visually impaired users is currently understudied. This paper aims to fill this gap.
Our study with 14 visually impaired users reveals that they are adapting these tools organically -- not only can these tools facilitate complex interactions in household, spatial, and social contexts, but they also act as an extension of users' cognition, as if the cognition were distributed in the visual information. We also found that although the tools are currently not goal-oriented, users accommodate this limitation and embrace the tools' capabilities for broader use. These findings enable us to envision design implications for creating more goal-oriented, real-time processing, and reliable AI-powered assistive technology. 
\end{abstract}


\section{Introduction}
Due to the absence of visual cues, people with visual impairments (PVI) encounter significant challenges in perceiving their surroundings. In recent years, PVI have relied on either AI-powered or human-assisted systems for visual interpretation and question answering to mitigate this challenge. Within the realm of AI-powered solutions ~\cite{ahmetovic2020recog, mukhiddinov2022automatic, hong2022blind, morrison2023understanding, gonzalez2024investigating}, advancements in computer vision (CV) and natural language processing (NLP) technologies, particularly the integration of deep learning models, have facilitated the identification of objects or text within scenes and the provision of responses to queries based on photos taken by PVI. In their daily routines, PVI frequently utilize commercial AI-powered applications such as Microsoft's Seeing AI~\cite{SeeingAI2020} to identify objects using device cameras.

More recently, Large Multimodal Model (LMM)~\cite{yu2023mm}, represented by GPT-4~\cite{achiam2023gpt}, have showcased remarkable proficiency in multimodal tasks like Visual Question Answering (VQA). Recognizing this significant advancement, researchers have begun preliminary investigations into the potential of LMM to support PVI~\cite{zhao2024vialm, yang2024viassist}. Leading this exploration is \bma~\cite{bma_usecase}, the first LMM-based system for visual interpretation and question answering designed to assist PVI. Powered by OpenAI's GPT-4 models~\cite{achiam2023gpt}, \bma{} offers superior capabilities compared to similar applications.

While prior studies~\cite{granquist2021evaluation, kupferstein2020understanding, gonzalez2024investigating} have examined the use cases of AI-powered visual assistive systems into PVI's daily lives, they have not incorporated the latest LMM technologies.
We investigate how LMM, specifically implemented in Be My AI, mitigate or address the limitations found in pre-LMM AI models, such as Microsoft’s Azure AI Vision image description API used in~\cite{gonzalez2024investigating} and Seeing AI. These pre-LMM AI models lacked the ability to provide comprehensive visual interpretations and satisfactory user engagement, failing to allow for follow-up questions or to deliver insightful opinions on visual content. Our work, distinct from earlier studies on pre-LMM AI models, is timely and responsive to the latest technological advancements. It offers unique insights into real-life effectiveness and user satisfaction with LMM-based VQA systems.

This study aims to explore how state-of-the-art AI opens up new opportunities for interactions through advanced human-like language and vision capabilities. We delve into the underexplored potential of LMM technologies in supporting PVI, going beyond the scope of previous AI efforts and focusing on the following \textit{research question}:
%
\begin{itemize}
  \item[\textbf{RQ.}] \textit{What are the emerging practices of LMM assistance in the daily lives of people with visual impairments?}  
\end{itemize}

To answer this question, we conducted an exploratory study by interviewing 14 users with visual impairments, who chose to adopt \bma{} into their daily lives and had been using it for at least two months. 
This approach allowed us to capture authentic user experiences with the LMM-based assistance in their real lives, rather than introducing a novel artifact that participants were unfamiliar with.


We found that Be My AI can enrich visual interpretations and integrate thoughtful opinions on visual content, facilitating new use cases and practices previously unachievable by pre-LMM AI models. For example, Be My AI has suggested recipes during cooking (Section~\ref{cooking}), offered fashion suggestions for creating style-coordinated outfits rather than merely identifying colors or patterns (Section~\ref{fashion}), verified images that participants planned to post on social media (Section~\ref{digital}), enriched social interactions with families (Section~\ref{physical}), and ensured safety in human-animal interactions (Section~\ref{animal}). However, it also presents challenges, including not always delivering goal-oriented information (Section~\ref{appliances},~\ref{navigation}, and~\ref{physical}) and occasionally producing AI hallucinations (Section~\ref{scene}).


Our findings suggest that \bma{} becomes part of the cognitive system in these contexts. 
We examine how participants offload visual cognition to \bma, accommodate its limitations, and creatively adapt its capabilities using their own cognition, such as memory, inferential reasoning, and experiential knowledge. 
This interaction between participants and \bma{} exemplifies the principles of distributed cognition, indicating a dynamic and collaborative cognitive relationship where visual perception is externalized to the artifact. 
We organized these insights into two categories in our findings: \textit{``Artifact as Cognitive Support''} and \textit{``Individual Adaptation and Cognitive Strategies''}. This template offers a structured approach to understanding how the tool becomes part of the cognition system.







This research provides insights into how advanced AI assistance, specifically LMM-based systems like \bma, are reshaping accessibility tools for PVI. By examining user interactions and adaptations, our study contributes to a deeper understanding of the cognitive dynamics involved in the use of such assistance. The findings not only highlight the practical applications of \bma{} but also pave the way for future innovations in assistive technology that can enhance autonomy and engagement for this community.

\section{Background and Related Work}

In this section, we review the literature on AI-powered and human-assisted visual interpretation and question answering systems, as well as distributed cognition.

\subsection{AI-Powered Visual Interpretation and Question Answering Systems}

The advancements in deep learning models for CV and NLP technologies have significantly enhanced the capabilities of AI-powered visual assistive systems. These systems leverage photos taken by PVI to identify objects or text within the scene, as well as respond to queries about the contents of the images. For instance, Ahmetovic et al.~\cite{ahmetovic2020recog} developed a mobile app utilizing deep network models to guide PVI in capturing images and recognizing objects. Mukhiddinov et al.~\cite{mukhiddinov2022automatic} devised a fire detection and notification app for PVI using convolutional neural networks. Hong et al.~\cite{hong2022blind} created an iOS app enabling PVI to gather training images for personalized object recognition. Morrison et al.~\cite{morrison2023understanding} designed an application to teach AI to identify individualized items, aiding PVI in locating personal belongings. Gonzalez et al.~\cite{gonzalez2024investigating} introduced a scene description app utilizing Microsoft's Azure AI Vision image description API. Additionally, PVI commonly employ commercial AI-powered applications like Microsoft's Seeing AI~\cite{SeeingAI2020} to recognize objects through device cameras.

More recently, Large Multimodal Model (LMM)~\cite{yu2023mm}, represented by GPT-4~\cite{achiam2023gpt}, have showcased remarkable proficiency in multimodal tasks. In response, researchers have begun exploring the potential of LMM in assisting PVI. Zhao et al.~\cite{zhao2024vialm} investigated the feasibility of leveraging state-of-the-art LMM to aid PVI and established a corresponding benchmark. Yang et al.~\cite{yang2024viassist} utilized LMM to develop an assistive system for answering PVI's questions about captured images, including assessing picture quality and suggesting retakes. In the realm of commercial applications, Be My Eyes~\cite{BeMyEyes2020} and OpenAI collaborated to introduce the \bma{} feature~\cite{bma_usecase}, powered by GPT-4~\cite{achiam2023gpt}, aimed at replacing human volunteers.

Prior works have delved into the utilization of AI-powered visual assistive systems by PVI in their daily lives. For instance, Granquist et al.~\cite{granquist2021evaluation} examined the use of Seeing AI for reading tasks among PVI participants, while Kupferstein et al.~\cite{kupferstein2020understanding} conducted a diary study to investigate PVI users' incorporation of Seeing AI into their daily routines. Recently, Gonzalez et al.~\cite{gonzalez2024investigating} undertook a diary study to explore the use cases of AI-powered scene description applications. However, these studies did not incorporate the latest LMM technologies, thus failing to represent the user experience of cutting-edge AI-powered systems. Bendel~\cite{bendel2024can} documented his experience with GPT-4-based \bma{}, yet his account remains subjective. Therefore, extensive and comprehensive research into PVI's daily utilization of state-of-the-art AI-powered systems is imperative. This paper addresses this gap by investigating the daily usage of \bma{} among fourteen PVI users.





\subsection{Human-Assisted Visual Interpretation and Question Answering Systems}

Human-assisted Visual Question Answering (VQA) systems offer crucial prosthetic support for PVI by facilitating connections to sighted people through remote assistance. These systems utilize image-based and video-based modalities to cater to different needs and situations.

Image-based human-assisted VQA systems allow PVI to submit photos along with their queries and receive responses after some time. An example of this is Vizwiz, where PVI can upload images accompanied by audio-recorded questions and receive text-based answers through crowdsourced assistance~\cite{bigham2010vizwiz_nearly}. This method has been successfully applied in tasks such as reading text, identifying colors, locating objects, and obtaining fashion advice~\cite{bigham2010vizwiz_nearly, bigham2010vizwiz, burton2012crowdsourcing}. Additionally, Vizwiz Social extends this functionality by linking PVI with their social networks, enabling support from friends and family members~\cite{gurari2018vizwiz}. However, the single-photo, single-query limitation of image-based VQA systems~\cite{bigham2010vizwiz_nearly} makes it less suitable for addressing complex or contextually deep inquiries~\cite{lasecki2013answering}.

Conversely, video-based human-assisted VQA systems facilitate real-time, interactive support, allowing PVI to receive immediate assistance tailored to their specific environmental context. This approach enables the visual interpretation of real-time scenes and supports a dynamic, back-and-forth VQA process, which is essential for addressing more specific and complex contextual inquiries effectively.
The evolution of this technology has progressed from wearable digital cameras ~\cite{hunaiti2006remote,garaj2003system,baranski2015field} and webcams~\cite{bujacz2008remote,scheggi2014remote,chaudary2017tele} to the utilization of mobile video applications~\cite{holmes2015iphone, BeMyEyes2020, Aira2020,xie2022dis,xie2024bubblecam} and real-time screen sharing technologies~\cite{lasecki2011real,lasecki2013answering,xie2023two}. 
Services like Be My Eyes~\cite{BeMyEyes2020}, which connects PVI with untrained volunteers, and Aira~\cite{Aira2020}, which connects PVI with trained professional assistants, exemplify the application of video-based VQA in scenarios that require immediate feedback. These services prove effective in navigation~\cite{kamikubo2020support,xie2022dis,c4vtochi,iui}, 
shopping~\cite{c4vtochi,xie2023two,iui}, 
and social interaction~\cite{lee2020emerging,Caroll2020Human,lee2018conversations}. 
In this study, we compare \bma{} with video-based human-assisted VQA systems to understand \bma's unique advantages and challenges.

\subsection{Distributed Cognition}


Distributed cognition is a framework that conceptualizes cognitive processes not as confined within an individual's mind, but rather as interactions among individuals, artifacts, and socio-cultural elements~\cite{hutchins1995cognition}.
In this framework, a cognitive process is defined by the functional relationships among the elements that participate in it, rather than by their spatial colocation~\cite{hollan2000distributed}. Furthermore, cognitive processes often require coordination between internal and external representations~\cite{rogers2022hci,hollan2000distributed} and can be distributed across time~\cite{hollan2000distributed}.

Research in HCI has explored the application of distributed cognition in explaining how people utilize external cognitive resources, like artifacts and environmental elements, to tackle complex challenges~\cite{halverson2002activity,baumer2011comparing}. 
For example, Wu et al.~\cite{wu2008collaborating} investigated how families dealing with memory impairments collaborate on everyday tasks, examining the mechanisms for accessing and sharing knowledge~\cite{rogers2006distributed}, facilitating communication and coordination within distributed cognitive framework~\cite{perry2003distributed}. 
Nobarany et al.~\cite{nobarany2012facilitating} applied this framework to explore how analysts, reasoning artifacts, and user interface elements contribute to supporting unanticipated reuse in distributed collaboration. 
In this study, we examine how PVI off-load cognitive efforts such as visual, remembering, and reasoning onto artifacts like \bma{} to achieve their goals. Additionally, we explore how these artifacts support and extend PVI's cognitive processes in their daily activities.

\section{Method}

We conducted interviews with $14$ visually impaired users to investigate the emerging practices of \bma{} in their daily lives. This study is IRB-approved.

\subsection{Participants}

We recruited a total of 14 visually impaired participants (4 males and 10 females, 10 blind and 4 low-vision) through our prior contacts and snowball sampling.  
Each visually impaired participant used and was familiar with Be My AI. 
Their common age groups are 35-40. Three of them are students, two of them are unemployed, and the rest were employed. 
Table~\ref{user_demographic_info} presents their demographics. 
Each visually impaired participant received a \$30 gift card per session for their time and effort.

\begin{table}[]
\small
\caption{Participants' demographics.}
\label{user_demographic_info}
\begin{tabular}{p{0.3cm}p{0.8cm}p{1cm}p{3cm}p{2.8cm}p{2.5cm}p{2.5cm}}
\toprule
\textbf{ID} & \textbf{Gender} & \textbf{Age Group} & \textbf{Condition of Vision Impairment} & \textbf{Age of Onset} & \textbf{Occupation Type} & \textbf{Be My AI Usage Frequency} \\ \toprule
P1 & F & 45-50 & Totally blind, retinopathy of prematurity & Since birth & IT consultant &  3 or 4 times a day \\ \hline
P2 & F & 35-40 & Low vision, cone-rod dystrophy & Since birth & Program director in a nonprofit &  a few times a week \\ \hline
P3 & F & 30-35 & Totally blind, Leber's congenital amaurosis & Since birth & Elementary school teacher & 5 times a week \\ \hline
P4 & M & 25-30 & Totally blind, Pale optic nerves & More than 12 yrs ago & Criminal law employee & 2 to 3 times a week \\ \hline
P5 & F & 40-45 & Totally blind, retinopathy of prematurity & Since birth & Manager of digital accessibility & 2 times a day \\ \hline
P6 & F & 25-30 & Totally blind, microcephaly and detached retina & Since birth & Student & once a week \\ \hline
P7 & F & 35-40 & Low vision, retinopathy of prematurity & Since birth & Part-time employee &  2 times a week \\ \hline
P8 & M & 40-45 & Totally blind, detached retina & Since birth & Insurance & a few times a day \\ \hline
P9 & F & 30-35 & Totally blind, retinopathy of prematurity & Since birth & In-between jobs & a few times a week \\ \hline
P10 & F & 30-35 & Low vision, retinitis pigmentosa & Since birth & Student &  3 or 4 times a day \\ \hline
P11 & M & 35-40 & Totally blind, retinopathy of prematurity & Since birth & Stay-at-home parent &  3 or 4 times a day \\ \hline
P12 & M & 20-25 & Low vision, Leber's hereditary optic neuropathy & Since 14 yrs old & Student &  2 times a week \\ \hline
P13 & F & 40-45 & Totally blind, retinitis pigmentosa & Low vision since infancy, totally blind since 2021 & Human service employee &  5 times a week \\ \hline
P14 & F & 35-40 & Totally blind, retinopathy of prematurity & Since a few months old & Assistive technology specialist &  5 times a week 
\\ \bottomrule
\end{tabular}
\end{table}

\subsection{Procedure}
We conducted 14 semi-structured interviews via Zoom, each ranging from 50 to 76 minutes in duration. All interviews were recorded after participants' consent. One or two researchers were present in each session.

Firstly, we invited participants to share their personal use cases for the \bma{} app. To aid in their recall and to enrich the discussion, we presented a list of common use cases provided by Be My Eyes~\cite{bma_usecase}, asking participants to identify any similar experiences they had encountered. Follow-up questions were posed to delve further into their experiences with each identified use case.

Secondly, we gathered participants' feedback regarding the quality of visual interpretations provided by \bma. This included evaluating the visual interpretations in terms of accuracy, level of detail, error, and the appropriateness of of how people’s identities were interpreted and described.

Thirdly, we explored participants' experiences with \bma{} in comparison to other assistive tools they use, including both human-assisted and AI-powered VQA systems. This comparison helped to contextualize the unique advantages and challenges of \bma{} relative to other technologies.

Finally, we collected participants' demographic information and inquired about their willingness to share copies of the visual interpretations from \bma{} with our research team for detailed analysis.

\subsection{Data Analysis}




We used a bottom-up approach in our qualitative data analysis. The first author performed inductive thematic analysis~\cite{braun2006using} on the transcribed interviews, used open coding to develop initial codes, and generated themes and subthemes through iterative collating and grouping. The themes and subthemes were reviewed and finalized during weekly meetings with all authors. 
Next, we present our findings.

\section{Findings}
\label{sec:findings}

Seven participants (P1, P6-8, P10, P12, P14) highlighted their use of \bma{} for reading tasks, which is a prevalent need within VQA systems~\cite{avila2016remote,lee2020emerging,BME_100ways}. Their activities spanned reading mail, newspapers, and labels. 
In this section, however, we aim to explore beyond basic reading functions, which are already well-addressed by existing technologies like OCR, to examine the more emerging and interactive practices introduced by LMM technologies. Specifically, we delve into how \bma{} supports managing personal and household tasks, understanding spatial environments, and engaging in social interactions. 

Our analysis reveals that \bma{} is part of the cognitive system in these contexts. In each use case, we examine how participants offload their cognition to \bma{} in \textit{``Artifact as Cognitive Support''}, as well as understand how participants accommodate challenges and limitations associated with \bma{} and how they creatively adapt \bma's capabilities to more creative and personalized use cases in \textit{``Individual Adaptation and Cognitive Strategies.''}






\subsection{Managing Personal and Household Tasks}



In this section, we explore how \bma{} facilitates participants in daily home activities, including using household appliances, cooking and dining, and fashion help. It highlights how \bma{} improves the practical aspects of home and personal life.

\subsubsection{Help with Household Appliances}
\label{appliances}
Participants engaged \bma{} for assistance with household appliances, categorically divided into grid keypad and rotary control appliances. These devices necessitate tactile interaction for operational inputs, requiring actions such as pressing buttons or turning dials for setting controls or making selections. 
%





\paragraph{Artifact as Cognitive Support:}


Five participants (P2, P5, P9, P11, P14) utilized \bma{} to navigate the grid layout of keypad panels on appliances, characterized by organized rows and columns. 
\bma{} reduces participants' cognitive load of visually searching the layout by providing structured verbal descriptions.
For example, in detailing a microwave's control panel, \bma{} begins by identifying the number of rows and columns, followed by employing directional (e.g., top, bottom, left, right, center) and sequential (e.g., second row, from left to right) terminology to guide users through the panel's arrangement. 



However, we also identify limitations in the \bma’s ability to support all cognitive needs effectively in helping with household appliances. Six participants noted its inadequacy in accurately describing rotary control appliances like washers (P1) and thermostats (P2, P4, P5, P13, P14). These appliances require \bma{} to interpret current settings as users adjust dials for time, mode, or temperature.

Critically, participants observed that \bma{} lacks a goal-oriented approach, often offering broad descriptions of visual elements without honing in on the user's specific objectives. P12 highlighted this issue, remarking, \textit{``Be My AI loves to make general descriptions, and it doesn't know what to focus on,''} leading to user frustration. For example, while \bma{} recognized a thermostat on the wall, it failed to convey essential details such as the current temperature setting or instructions for temperature adjustment. 


\paragraph{Individual Adaptation and Cognitive Strategies:}
To overcome these limitations, participants developed adaptive strategies. For example, P5 posed targeted questions to \bma, such as \textit{``What is the arrow pointed at right now on the current setting?''} This approach refined \bma's output to be more precise and relevant to the setting, enabling P5 to adjust the dial accurately.
P12, on the other hand, aimed to improve \bma's focus by carefully aligning the camera with the specific item of interest, thus directing \bma{} to concentrate on particular elements rather than offering broad, scene-wide descriptions.
These adaptations indicate how users actively reshape their interactions to optimize cognitive support. 


Additionally, eight participants (P1-3, P7, P8, P12-14) compared \bma{} with video-based human-assisted VQA systems, noting that the latter offers more adaptive and contextually aware support in tasks that require continuous real-time feedback, such as dial adjustments.
P13 shared an instance where volunteers from \bme{} understand her intent beyond mere identification of objects like a thermostat.
Volunteers can provide immediate, actionable feedback and guidance, facilitating a more effective interaction through conversation.

\begin{quote}
    \textit{``I ended up pulling up a volunteer, and the volunteer was fantastic. He goes, `You know, these are tricky.' And he says, `It looks like it's set at 69. What do you want?' And I said, `71.' He goes, `Turn it, turn it. Yep, it's 71 from what I can tell.' And I went, `Cool.'''} (P13)
\end{quote}

In summary, \bma{} reduces the visual load involved in using household appliances, but it is not goal-oriented. Participants adapt \bma's capabilities to their intent by asking specific questions about the visual content or by seeking alternative means, such as calling human volunteers, to fulfill their goals.

\subsubsection{Cooking and Dining}
\label{cooking}
Participants leveraged \bma{} for various culinary tasks, including identifying specific foods among similar items, like drink cans (P6), coffee beans (P7), cans of tomatoes or avocados (P10), chili spices (P10), and Campbell's soup cans (P13). They also used it to access cooking instructions regarding time and temperature (P7, P8, P9), read non-Braille recipe books (P9), and identify ingredients and suggest recipe (P12).
%



\paragraph{Artifact as Cognitive Support:}

\bma{} transforms complex visual information, such as graphics on food packaging, into comprehensible audible text that PVI can process. It acts as a cognitive proxy, interpreting and relaying information that would otherwise be inaccessible, thus reducing participants' cognitive load associated with tasks such as identifying specific food items or following cooking instructions.

Unlike other AI-powered VQA systems like Seeing AI and OCR, \bma{} processes graphic elements in addition to text. 
For example, \bma{} informed P2 about an \textit{``allergy information emblem''} on product packaging, aiding in selecting allergy-safe foods. \bma{} detailed not just the brand name but also specific allergy-free labels that indicated the absence of common allergens like gluten, dairy, and sesame.


Moreover, \bma{} enhances the cooking and dining experience through its ``ask more'' function, which allows participants to interactively engage with the artifact to receive tailored cooking advice. P12 utilized this feature, who sought recipe suggestions by inputting available ingredients: \textit{``I typed in the chat that I have chicken, pasta sauce, onion... and I asked for \bma, you know, to provide a recipe.''}
%
Responding to P12's input, \bma{} offered practical cooking instructions that included a list of ingredients, preparation methods (e.g., boiling, dicing, cutting into bite-sized pieces), required cookware, steps detailed with timing and sequencing, and seasoning tips.



\paragraph{Individual Adaptation and Cognitive Strategies:}

Participants employed specific strategies to manage the cognitive load involved in cooking tasks. By leveraging \bma{} as an external memory aid, P12 accessed complex cooking instructions in a step-by-step manner, reducing the need to retain and remember extensive procedural details from cookbooks or websites. 
%
%
This interaction with \bma{} involved retrieving information, processing it, and then executing based on the guidance received, illustrating the dynamics between P12's cognitive processes and the use of the artifact. 

In summary, \bma{} converts the graphical content on food packing to audible formats and offers recipe recommendations for cooking and dining. Participants use \bma{} as a cognitive aid in accessing visual information and navigating the detailed recipes it provides.

\subsubsection{Fashion Help}
\label{fashion}
\bma{} has served as a valuable tool for fashion-related tasks, assisting participants in several key areas: identifying colors of clothes (P2, P7, P9, P13), detailing patterns or designs on clothes (P2, P9), offering suggestions for assembling stylistically harmonious and color-coordinated ensembles like outfits, shoes, and jewelry (P2, P4, P5, P6, P12), and checking makeup (P5, P10).
%



\paragraph{Artifact as Cognitive Support:}

Participants have found \bma{} exceptionally precise in taking the visual load by identifying colors and patterns on clothing, as well as in assessing makeup, with no discrepancies reported. For instance, P5 effectively used \bma{} to verify the color, placement, and overall balance of her makeup, ensuring there were no mismatches or issues.

This accuracy is particularly valued in fashion-related tasks, where \bma{} demonstrates a more consistent and reliable perception of color compared to human assistance. 
P4 pointed out that human assistants, such as \bme{} volunteers, often have subjective interpretations, stating, \textit{``each volunteer has a different way of seeing things''}. 
P9 echoed that \bma{} maintains consistent detail in its interpretations, whereas human volunteers can vary significantly in their visual descriptions: \textit{``Sometimes they might not understand exactly, or like how much you want it to be described; they don't tell you the same description.''}

P12's experience further underscores the reliability of \bma{} over human volunteers in color identification. He shared an instance where a colorblind volunteer, without disclosing the colorblindness, provided incorrect color information of a tie. 
This example underscores \bma's utility as a reliable tool for accurate color matching, essential for appropriate fashion decisions, as P12 mentioned: \textit{``This is a really good application for \bma{} because you don't have to rely on someone's color vision.''}


Moreover, \bma{} enhances decision-making in fashion by providing suggestions to create stylistically coordinated outfits. P9 consulted \bma{} to harmonize colors of a top and dressy pants effectively, and P12 used it to select a tie that complemented his shirt. 
A notable example is P2's use of the ``ask more'' function to complete a fall-season look with a turquoise top.  
Concerned about the seasonality of a turquoise top, P2 asked \bma, \textit{``Is this too summery of a color? Could I wear it in the fall?''} \bma{} suggested, \textit{``Turquoise is typically a bright color that's considered more of a summer-spring color, but you could tone it down into more of a fall look with a cardigan or chunky boots.''} Following this advice, P2 selected a grayish taupe pair of ankle boots, and upon capturing their image, \bma{} confirmed, \textit{``This might pair well with the turquoise top to make it more fall appropriate.''}

\paragraph{Individual Adaptation and Cognitive Strategies:}


Participants were divided regarding \bma's fashion suggestion functionality. Some appreciated this feature because it \textit{``definitely creates more confidence''} (P2) in preparing outfits for specific occasions, such as work or sports events. For instance, P5 found \bma{} helpful to \textit{``lay out [her] outfit before the race,''} enhancing her preparation and confidence for race day.

However, other participants (P6, P7) held the opposite views. 
While they utilized \bma{} to obtain descriptions of colors and patterns, they hesitated to rely on its fashion suggestion. These participants preferred to make their own judgment and choices about outfit matching, emphasizing the importance of human subjectivity in fashion decisions. 
P6 expressed concerns about the AI's ability to replicate human subjectivity, remarking, \textit{``It's interesting how AI is being taught to simulate kind of the human factor of things.''}
She linked her skepticism to instances where AI-generated responses were \textit{``strange''} and \textit{``complete nonsense,''} which contrasted with human creativity and nuanced understanding.

\begin{quote}
    \textit{``No, no, no, no, I would never use it to do anything that required human subjectivity... I just don't trust AI with a task that is supposed to be subjective like that, particularly visual like that. Have you ever seen AI weirdness?... I think that just goes to show why I'm not gonna trust AI with my fashion yet.''} (P6)
\end{quote}


In summary, \bma{} enhances decision-making in fashion by offering precise color identification and tailored fashion suggestions. Participants respond diversely to these capabilities. Some embrace the suggestions, finding that they bolster confidence and assist in assembling outfits, while others prefer to rely on their own judgment and decision-making to coordinate their attire.

\subsection{Understanding Spatial Environments}

In this section, we examine how \bma{} enhances spatial awareness through scene descriptions, locates dropped objects, and facilitates navigation. It highlights \bma's role in improving participants' ability to interact with and navigate their surroundings.

\subsubsection{Scene Description}
\label{scene}

Participants utilized \bma{} for detailed scene descriptions across both indoor environments, such as theaters (P2) and room layouts (P3, P7, P9, P13), and outdoor settings, including holiday decorations (P5) and street scenes (P11, P12). 
%

\paragraph{Artifact as Cognitive Support:}

Six participants (P2, P3, P7, P9, P13, P14) valued \bma's detailed scene descriptions for increasing their spatial awareness. 
It often reveals spatial details that are not initially apparent, as noted by P11, \textit{``It gives me unexpected information about things that I didn't even know were there.''} 
\bma{} provides comprehensive visual information, augmenting participants' visual cognition by detailing objects by color, size, shape, and spatial orientation (e.g., ``on the left,'' ``to the right,'' ``in the middle,'' ``in the background''). It also describes dynamic elements like bystander actions (e.g., ``sitting''), text on signs or advertisements, and the overall atmosphere (e.g., ``peaceful and natural,'' ``cozy and cheerful holiday vibe,'' ``warm and inviting atmoshpere'').

These detailed scene descriptions offers a richer sensory experience than those often provided by human assistants. Participants particularly appreciated \bma{} for its ability to handle visually-overloaded scenes with less subjectivity than human assistants. P2 pointed out that human assistants often feel overwhelmed by the abundance of visual elements, struggling to decide what to describe. 
In these scenarios, human assistants prioritize and focus on elements they deem important and potentially overlook other details. In contrast, P9 highlighted that \bma{} does not ignore details that human assistants might consider unimportant. 


\begin{quote}
    \textit{``Maybe a family member wouldn't think that I would want to know that there was like a picture frame hanging in the wall or something... But \bma{} just described everything, so like, it is nice because you get to imagine how something really might look or how the house might be decorated or things like that.''} (P9)
\end{quote}

Despite these advantages, there are notable drawbacks with \bma{} related to AI hallucinations, errors where the artifact inaccurately identifies objects that aren't present. 
For instance, \bma{} erroneously added nonexistent details to scenes, such as an object mistakenly reported behind P3 and incorrect features in P5's home. 
%


\paragraph{Individual Adaptation and Cognitive Strategies:}




To overcome issues with AI hallucinations, participants adopt various strategies to ensure accuracy in the visual information provided by \bma. For instance, P3 recounted an incident where \bma{} inaccurately reported an object behind her. To confirm this, P3 consulted a human assistant, who verified that there was indeed nothing behind her in the picture. 

On the other hand, P5 shared an experience where \bma{} added non-existent details to an image at her home, including elements supposedly alongside their dogs. Relying on her familiarity with the environment, P5 was able to identify these inaccuracies without external assistance. This reliance on personal knowledge illustrates how participants used their understanding of familiar surroundings to challenge and correct the AI's interpretations.





In summary, \bma{} acts as a cognitive extension by providing comprehensive scene descriptions that enhance participants' perception of their surroundings. Despite its efficacy, it is subject to the limitations of AI hallucinations. To overcome this issue, participants either consult human assistants for verification or rely on their own knowledge to pinpoint inaccuracies.


\subsubsection{Locating Dropped Objects}
\label{locate_object}
The sequential descriptions provided by \bma{} assist participants in locating dropped objects (P4, P12, P13) such as earbuds and hair ties. 

\paragraph{Artifact as Cognitive Support:}
Participants found that \bma's structured approach to describing surroundings aids in locating dropped objects. By systematically detailing the environment using a \textit{``top to bottom, left to right''} sequence, the artifact helps orient users more effectively compared to random starting points. 
For example, \bma{} enhances spatial orientation by providing specific relative locations. After a general description of the scene, it indicates, \textit{``picture of a carpet with a hair tie in the upper right hand corner,''} (P13) or notes 
\textit{``the earphones are directly in front of you, between your feet''} (P12). 
These descriptions allow participants to pinpoint lost objects with greater accuracy.

\paragraph{Individual Adaptation and Cognitive Strategies:}

Participants utilized \bma{} by integrating it with their own sensory and cognitive strategies to locate dropped objects. They enhanced their use of the artifact by relying on auditory cues and spatial memory, which helped in initially estimating the dropped object's location. This approach allowed them to optimize the camera angle for taking pictures that \bma{} would analyze. 
For instance, P4 used his \textit{``listening skills''} to determine the location where the object fell and guided where to point the camera accordingly. 
Similarly, P12 adjusted his position slightly backward from the seating area to capture a better overview of the floor where the earbuds were dropped. 
These adaptions illustrate how participants combine their understanding of the environment with \bma's technological capabilities to manage tasks that require spatial awareness.

In summary, \bma's sequential descriptions support participants in finding dropped objects. Participants leveraged their auditory skills and spatial memory to adjust the camera's position, allowing for more precise scanning and enabling \bma{} to focus on the intended search area.

\subsubsection{Navigation}
\label{navigation}
\bma{} is used for navigation tasks, such as localizing and orientating participants by reading airport signages (P2, P5), and store names when disoriented (P10). 
%

\paragraph{Artifact as Cognitive Support:}

Participants (P2, P5, P10) employed \bma{} as a tool to assist in localization and orientation when approaching their destinations. 
It reduces the cognitive process by offloading visual tasks such as reading signages and recognizing surroundings.
For instance, P2 utilized \bma{} to identify gate numbers at the airport, which helped her determine the location and navigate accordingly. 
Likewise, P5 used \bma{} to read a signage indicating the direction to the transportation area at the airport, while P10 took pictures of her surroundings for orientation when lost in her neighborhood.  
Such instances illustrate \bma's role in providing participants with \textit{``better spatial awareness of what's going on around [them]''} (P5), thereby allowing participants to gain \textit{``more independence and be more familiar with [their] environment''} (P10).

%

However, other participants (P6, P11, P13, P14) reported challenges with \bma{} for navigation, particularly due to its limited camera view and the practical issues of mobility. 
As P6 described, participants \textit{``have to stand there and keep taking pictures and taking pictures,''} check what's captured, determine if it's helpful for navigation, and adjust the angle for another pictures. This iterative process can be time-consuming and might cause self-consciousness, as \textit{``taking constant pictures with their camera could also look weird''} (P6) to bystanders.

Moreover, P13 critiqued that \bma{} is not goal-oriented enough to aid navigation because of its broad descriptions. For instance, she used \bma{} to find a specific office, but it failed to read the office's door number and instead \textit{``It sa[id] something like `wall with door,' which is, you know, not particularly helpful.''}


\paragraph{Individual Adaptation and Cognitive Strategies:}

In navigation, participants combined \bma's visual cognition with other sources of guidance. When using \bma, they often knew they were near their intended areas based on prior help from human assistants (P2, P5) or through their existing mental map (P10). For instance, after getting directions from a human assistant and was seated in the Gate 12 area, P2 leveraged \bma{} to move from Gate 12 to Gate 13. This demonstrates the integration of multiple sources of information.

Despite the varied feedback on \bma, participants agreed that \bma{} is \textit{``not a replacement for [their] mobility skills or just any skills in general''} (P14). P5 elaborated that due to the limited information captured and interpreted by \bma, it can only provide partial navigational details, such as the presence of obstacles, noting \textit{``it's not gonna be able to tell you, like, to protect you every step of the way.''}

Consequently, participants integrate \bma{} with their orientation and mobility (O\&M) skills to enhance their navigation. P13 mentioned that while \bma{} can indicate the presence of obstacles like stairs, it does not provide essential details such as the distance to these obstacles or their characteristics, which are typically discerned through O\&M tools like white canes or guide dogs.

\begin{quote}
    \textit{``[\bma] says, you know, `stairs in front,' and it's like, `Okay, that's great, but where are they? How far are they? Are they going up? Are they going down? Is there a railing?' which would be information that the dog or the cane could tell you. So, I would say use it as a tool along with, but definitely not by itself.''} (P13)
\end{quote}

In summary, \bma{} boosts spatial awareness and independence, although some participants question its practicality and safety during navigation. However, there is consensus that \bma{} can only complement, rather than replace participants' O\&M skills.

\subsection{Engaging in Social Interactions}

In this section, we assess how \bma{} supports various social engagements, including digital and physical human interactions, as well as human-animal interactions. It highlights \bma's role in enhancing users' social experiences across different settings.

\subsubsection{Digital Human Social Interactions}
\label{digital}

Participants leveraged \bma{} to engage in social interactions with blind and sighted people on social media by using it to describe pictures that are posted on social media (P1-4, P7, P10, P11), check the quality of pictures that participants intend to post on social media (P2, P4, P9), and generate descriptions of their own pictures to post on social media (P2, P3, P7).

\paragraph{Artifact as Cognitive Support:}
Pictures posted on social media often lack alt text, which describes an image's content and is accessible through screen readers.
Seven participants (P1-4, P7, P10, P11) value \bma's detailed interpretations of images posted on social media that are missing alt text, allowing them to fully comprehend the visual content. For example, P7 illustrated the effectiveness of \bma{} in contrast to Facebook's AI image description tool. 
While Facebook AI might provide a vague description such as \textit{``picture of five people,''} \bma{} provides a vivid depiction, such as describing a scene with someone skiing on a mountain alongside a visible ski lift. 
This level of detail enhanced P7's immersive experience, 
as she put it, \textit{``[\bma] gives me a clearer picture of what's around, especially in pictures of, you know, the scene. I can visualize it as to the point where I'm almost there.''}


Furthermore, \bma{} not only assists participants as recipients but also empowers them to become active contributors, enhancing their engagement on social media platforms. First, \bma{} supports participants in ensuring the quality and appropriateness of images they intend to post on social media (P2, P4, P9). It helps them check content clarity and composition to select suitable images. 
Second, \bma{} assists in creating accessible alt text (P2, P3, P7), facilitating easier access for other visually impaired users on social media. 
It allows participants to add personalized details such as familiar names or specific characteristics. 

\paragraph{Individual Adaptation and Cognitive Strategies:}
The adoption of \bma{} has altered the social media behavior of participants, enhancing their confidence and engagement. By understanding the visual content of images, participants like P10 are more confident in posting pictures, overcoming previous hesitations due to uncertainty about the visual aspects of their posts: \textit{``Before, I was very shy to post pictures because I didn't know the visual description of them.''} 
This tool also enables them to engage more actively with social media content. For instance, they can now provide detailed comments on friends' photos (P7) and use the content as a basis for conversations (P1, P3), achieving a level of social interaction similar to that of sighted users. 



Moreover, \bma's accurate image descriptions also affect the behaviors of sighted individuals who share images with participants. It alleviates the pressure on sighted people to provide perfect descriptions, thereby facilitating more frequent and meaningful interactions. As noted by P3, \textit{``Sighted people who now feel like they can send pictures without the pressure of having to describe it perfectly,''} indicating that \bma{} helps bridge the communication gap between visually impaired and sighted communities.

In summary, \bma{} provides detailed image descriptions for participants, enabling them to access visual content on social media platforms. This capability allowed visually impaired and sighted users to interact more confidently and effectively in digital human interactions, promoting enhanced social engagement.

\subsubsection{Physical Human Social Interactions}
\label{physical}

\bma{} has been utilized for parenting (P4, P8, P9), facilitating gift exchanges (P2, P9, P12), and helping identify puzzles to play with families (P14).

\paragraph{Artifact as Cognitive Support:}
In their parenting roles, participants (P4, P8, P9) effectively engaged \bma{} to offload the cognitive tasks of visual perception. They utilized \bma{} to describe and verify the appearance of their children, focusing on attributes like hair texture, skin complexion, and the styles and colors of clothing. This enabled them to dress their children appropriately. 
Similarly, P9 employed \bma{} to read and label books in Braille, enhancing her ability to contextualize stories and interact more engagingly with her child.

\bma{} also facilitates enhanced engagement within family settings by providing detailed descriptions. P14 utilized \bma{} to differentiate between various puzzles by describing the images on the boxes, such as \textit{``whether it was a puzzle, a scene of cats, or a scene of bears.''} This capability is particularly notable when compared to human assistants, as P14 pointed out that \bma{} offers access to visual details that sighted family members often take for granted. For example, \bma{} read the text from the box specifying which piece needed to be placed in the center, a detail not initially provided by human assistants. In this context, \bma{} offloads the cognitive load associated with visual distinction, enabling P14 to participate more fully in family activities.


Despite its advantages, \bma{} has certain limitations that can hinder its effectiveness in providing goal-oriented information. P14's experience with a puzzle box illustrates this issue. While \bma{} could identify general shapes, it sometimes lacked the specificity required for more nuanced tasks.
%

\begin{quote}
    \textit{``When \bma{} read me the text of the box on the back of the puzzle that said, you know, this particular piece should be in the center of the puzzle. As a follow-up question I asked, `Could I have more information about the piece in the center?' And it said, `This piece is a square piece,' but I mean, there were many different square pieces, so I could not tell from that.''} (P14)
\end{quote}
This example shows that although \bma{} can process visual information, it may not always provide the level of detail needed to accomplish participants' goals.

\paragraph{Individual Adaptation and Cognitive Strategies:}

The detailed visual interpretation capability of \bma{} enables participants to engage in more meaningful social interactions.
This adaptation was particularly evident in their approach to gift exchanges, which became more personalized and thoughtful. For example, P2 ensured that the price tags were removed from Christmas gifts before wrapping them, and P9 used \bma{} to read party invitations and prepare gifts that matched the theme. 


P12 used \bma{} to describe Christmas gifts, explaining that reading cards with \bma{} \textit{``makes it a more meaningful experience with other people.''}
P12's experience illustrates how \bma{} enhances individual perception and, consequently, genuine social engagement. Before using \bma, opening Christmas cards was a performative act—pretending happiness despite not being able to see or understand the card's content. 
With \bma, P12 could capture the card's visual and textual details, allowing for authentic reactions to the messages. 


In addition to enhancing individual perception, \bma{} plays a role in decision-making within physical social engagement. P14 highlighted this when describing her use of \bma{} to select a puzzle:
\textit{``I wanted part of my decision of putting a puzzle together to be, `Let me decide from a description of what the puzzles are that somebody could see on the box.'''} 
P14 stressed the importance of this autonomy, noting that her decision to engage in the puzzle activity was enriched by having access to the same visual information that sighted people see on the box.



In summary, \bma{} provided detailed visual interpretations that enabled participants to perform tasks independently and interact more authentically in physical social interactions. For example, participants used \bma{} to enhance engagements in family activities through parenting and gift exchanging. They also used \bma{} to facilitate decision-making and autonomy in choosing puzzles.

\subsubsection{Human-Animal Interactions}
\label{animal}

Nine participants (P2, P4, P5, P8-11, P13, P14) utilized \bma{} to take leisure pictures of animals, including cats, dogs, birds, and horses, to know their status.

\paragraph{Artifact as Cognitive Support:}
Participants employed \bma{} to enhance their perception of animals' states and behaviors, which are critical aspects of interacting safely and effectively with animals. \bma{} acted as an external cognitive aid that enabled participants to access visual information. 
For example, participants (P5, P10, P11) used \bma{} to grasp nuances such as animals’ facial expressions, activities, and body language. P10, in particular, highlighted the utility of \bma{} during walking her dog, where she put it, \textit{``Sometimes it's hard for me to know if the dog is peeing or what the dog is doing.''}

Furthermore, P11's application of \bma{} in a farm setting exemplifies its role in enhancing awareness and safety for human-animal interactions. He uses the artifact to determine the locations and directions of mini horses in stalls for safety measure, like \textit{``Is the horse's head facing towards me or away from me? Is the horse turned towards the side? Based on where I am, where is its tail or hind, or its back end located?''}
P11 shared an incident where he was between two horses and was accidentally kicked by a mini horse that reacted jealously to his attention towards another larger one. 
This incident underscores the necessity of \bma{} in preempting dangerous situations, thus allowing P11 to take preventive measures. For example, using lead rope to \textit{``turn [the mini horse's] head the other way, and change her direction, get her focus back on me so she wouldn't have kicked me.''}


\paragraph{Individual Adaptation and Cognitive Strategies:}

While \bma{} provides valuable information in human-animal interactions, participants often adapted the tool's outputs to align with their own perceptions and combine them with other sensory information to ensure accuracy and safety.

P5, for example, emphasized personal judgment over the tool's subjective interpretations. She preferred the flexibility to override \bma's assumptions about her dogs' expressions. This strategy allowed P5 to balance AI-based interpretations with her subjective inputs, reinforcing her sense of control in understanding pet behaviors.
\begin{quote}
    \textit{``I like having more control over the description. So I like the ability to edit that. Some blind people think, `How does it know that the dogs are happy? Why does it assume?' Some people don't like that it's making assumptions about the picture. I like having access to that information, but I like to be able to change it if I want.''} (P5)
\end{quote}


P11 integrated visual interpretations from \bma{} with auditory, tactile, and olfactory cues to bolster safety during interactions with horses. He supplemented this information by consulting AI tools like ChatGPT, which guided him on how to interact safely based on sensory cues. For instance, he asked for advice on how to handle horses safely without sight, such as maintaining close contact and physical awareness of the horse's movements. This approach demonstrates how P11 skillfully combined technological insights with his sensory perceptions to ensure safety and navigate his interactions with horses effectively.

In summary, participants delegated the visual perception of monitoring the animal status to \bma. They further adapted \bma's output by incorporating their own perceptions and other sensory information to ensure accuracy and safety.





\section{Discussion}
\label{sec:discussion}


In this section, we delve into how participants and \bma{} collaborate as interconnected components within a distributed cognition framework. 
We examine how participants offload the task of visual perception to \bma, transforming it into a cognitive prosthetic that enhances their ability to navigate and interact with their environment. 
%
Additionally, we discuss design implications for AI-powered assistance.

\subsection{Task Offloading to \bma{} Through Distributed Cognition}







Our findings reveal that participants and \bma{} collaboratively work as interconnected components within the cognitive system, reflecting the principles of distributed cognition. Participants offload the burden of visual perception to \bma, enabling them to more effectively navigate and interact with their environment. This tool acts as a prosthetic for visual processing, transforming complex visual information into comprehensible audible text, which PVI can easily process.

Unlike traditional prosthetics like the white cane, which extend PVI’s tactile senses, \bma{} is an active cognitive extension. It interprets visual data in real time, providing dynamic and creative assistance in various contexts. 
For example, it can read allergy-free labels on food packaging (Section~\ref{cooking}), identify colors and patterns for outfit matching (Section~\ref{fashion}), and describe scenes to enhance spatial awareness (Section~\ref{scene}). 
With its capability to interpret visual content in a detailed and accessible manner, \bma{} effectively simulates visual cognition for PVI.

Despite its advantages, \bma{} also presents several limitations and challenges.
To overcome these obstacles, participants have developed adaptive strategies.
They utilize their memory, inferential reasoning, and experiential knowledge alongside \bma{} to create a more robust cognitive extension within the distributed cognition system. 
For instance, when \bma{} recognized a thermostat but failed to give detailed information about how to adjust a thermostat, participants posed targeted questions to get more specific information or called for human assistance (Section~\ref{cooking}). 
In cases of AI hallucinations, participants rely on their spatial knowledge or seek human confirmation (Section~\ref{scene}). These adaptations illustrate the dynamic nature in distributed cognition, where participants actively reshape their interactions with technology to enhance cognitive support.
%

Furthermore, participants creatively adapt \bma's visual interpretation capabilities to their personal needs and contexts. For instance, they combine \bma's outputs with their O\&M skills to enhance navigation (Section~\ref{navigation}) and supplement \bma's visual cues with their auditory, tactile, and olfactory senses to ensure safety when interacting with horses (Section~\ref{animal}). 
Participants also adjust their use of \bma{} when assessing fashion choices (Section~\ref{fashion}) or interpreting a pet’s expressions (Section~\ref{animal}).

These examples demonstrate how users actively shape their cognitive support systems, reinforcing the idea that distributed cognition is a dynamic process. By harnessing their own cognitive strengths and adjusting their interaction patterns with \bma, participants create a more flexible and robust approach to accessing visual information.








\paragraph{Comparison with Non-LMM AI-Powered Systems}

Compared to non-LMM AI-powered systems like Seeing AI~\cite{SeeingAI2020} and Facebook AI, participants noted that \bma{} can process both graphic elements and text, offering more comprehensive and vivid interpretations of visual content (Section~\ref{cooking}). This capability enhances the distributed cognition system by incorporating a form of visual cognition that is superior to conventional systems.

Additionally, \bma{} facilitates interactive conversations with its users, a functionality absent in OCR-based visual interpretation systems. This conversational ability introduces an interactive cognitive process, where \bma{} not only decodes visual information but also engages in a dialogic interaction. This interaction enhances cognitive engagement, allowing users to clarify, question, and explore visual content in a more dynamic manner.



\paragraph{Comparison with Human-Assisted VQA Systems}
In human-assisted VQA systems, another human’s cognition is involved to support PVI. In these systems, participants offload visual processing and some aspects of reasoning to human assistants. 
For instance, human assistants can identify the colors of outfits; however, their capability may be limited by factors such as colorblindness (Section~\ref{fashion}) or they may get overwhelmed by an abundance of visual elements and overlook critical details (Sections~\ref{scene} and \ref{physical}). Moreover, services like Aira instruct their human assistants to provide only objective information~\cite{lee2020emerging,xie2023two}, which may limit the depth of cognitive support provided. 

In contrast, \bma{} offers consistent visual interpretations and goes beyond mere description by incorporating subjective judgments, like fashion suggestions. 
This capability allows \bma{} to not only support basic visual tasks but also integrate contextual understanding into its interactions. 
By doing so, \bma{} functions as an extension of the PVI's cognitive processes, adapting to various contexts and individual needs in ways that human assistants may not. Such a system enriches the distributed cognition framework by providing a more adaptive cognitive agent. This adaptability enables PVI to engage with their environments more seamlessly.

\subsection{Design Implications for LMM Assistance for People with Visual Impairments} 
As demonstrated by our research, \bma{} has shown significant potential in empowering PVI by providing a more intuitive, user-friendly, and context-aware assistive experience. However, to further enhance the usability and effectiveness of AI-powered VQA systems like \bma{}, several design implications should be considered.
%
%

\subsubsection{Towards Goal-Oriented AI-Powered Visual Assistance}

%

Our findings highlight a limitation in the capabilities of \bma{} when it comes to providing actionable, goal-oriented guidance to PVI (Section~\ref{appliances}). While \bma{} is good at conveying ``what'' information with most of time accurately describing the visual content of a scene or object, such as identifying a thermostat on the wall. 
It often struggles with providing ``how'' information, guiding users on the specific actions required to interact with or operate elements in their environment, such as how to adjust the theromstat. This ``what'' and ``how'' divide poses a major challenge to the effectiveness and usability of \bma{}, as PVI rely on it not only for understanding their surroundings but also for completing tasks and achieving their goals.

To address this limitation and design more goal-oriented AI-powered assistance, we propose the following key design implications. 
Future AI-powered assistive technologies should be designed with a focus on action-oriented reasoning and task-specific guidance. Although this could be achieved through further user inquiries~\cite{truhn2023large}, integrating knowledge bases~\cite{zhu2014reasoning} and event/behavior reasoning engines~\cite{chen2008using} to enable contextual inference of actions and intentions, and associating visual elements' feedback with reasoning, would greatly reduce the cognitive burden on PVI and enhance the user experience. By leveraging this knowledge, assistive technologies can provide more relevant and actionable guidance to PVI, helping them effectively navigate and interact with their environment to complete desired tasks. Our findings emphasize the importance of human-centered design principles, particularly in the design of assistive technologies, which should be reinforced through a goal-oriented technical roadmap that adapts to users' needs, preferences, and external environments~\cite{fischer2001user,amershi2014power}. By emphasizing action goal-oriented reasoning~\cite{huffman1993goal,letier2002agent}, future AI-powered assistance will be optimized, further benefiting PVI.

\subsubsection{Towards Real-Time Processing AI-Powered Visual Assistance}
One of the most significant advantages of \bma{} and other LMM-based assistive tools is their ability to provide contextually relevant and personalized assistance to users. By leveraging the power of machine learning and the ability for understanding of natural language, these systems can understand and respond to a wide range of user queries. This level of contextual awareness represents a significant advancement over traditional assistive technologies, which often struggle to adapt to the diverse needs and preferences of individual users.

However, our findings also identify several challenges and limitations of current LMM-based assistive tools, particularly in terms of their reliance on user-generated images. Participants in our study reported frustration with the need to take multiple pictures to capture the desired information, which can be time-consuming and cognitively demanding (Section~\ref{navigation}). To address this issue, we envision the integration of real-time image processing capabilities into future AI-powered VQA systems. By continuously analyzing the user's surroundings and providing relevant information without the need for explicit image capture, these systems could offer a more seamless and efficient user experience. The integration of real-time image processing into AI-powered VQA systems aligns with the growing availability of commercial products and research prototypes that leverage advanced object detection and text recognition technologies. For example, Microsoft's Seeing AI \cite{SeeingAI2020} and various currency recognition systems \cite{liu2008camera, parlouar2009assistive, paisios2012exchanging} demonstrate the feasibility and potential impact of real-time image processing in assistive technology. By building upon these existing approaches and incorporating state-of-the-art deep learning techniques for object detection \cite{girshick2014rich, girshick2015fast, ren2016faster, krizhevsky2017imagenet} and text recognition \cite{ma2018arbitrary, he2017deep, zhou2017east, yao2016scene, liu2017deep, lyu2018multi}, future AI-powered VQA systems could provide even more robust and reliable assistance to PVI.

\subsubsection{Towards Reliable AI-Powered Visual Assistance}

Our findings highlight the significant potential of AI-Powered assistive technologies like \bma{} in enhancing the perception and understanding of surroundings for PVI. However, our study also reveals a notable drawback of AI-powered visual assistance, namely AI hallucinations. These errors, where the artifact inaccurately identifies objects that aren't present, can lead to confusion and mistrust among users. Participants in our study reported instances where \bma{} erroneously added non-existent details to scenes (Section~\ref{scene}). 
It can be argued that the presence of AI hallucinations poses a major challenge to the reliability and robustness of AI-powered assistance. If users cannot trust the information provided by these systems, their effectiveness as cognitive extensions is severely compromised. This issue is particularly critical for PVI, who rely on these technologies to navigate and make sense of their environment.

To address the problem of AI hallucinations, our participants applied various strategies, such as consulting human assistants for verification or relying on their own knowledge to identify inaccuracies. While these strategies showcase PVI's adaptability and problem-solving skills, they also highlight the need for more reliable and robust AI-powered assistive technologies.

One potential approach to mitigating AI hallucinations is to incorporate uncertainty estimation and communication into the design of these technologies. By quantifying and conveying the confidence level of predictions, AI-powered assistive systems can help users assess the reliability of the information provided. This approach has been explored in the context of other AI-based systems, such as medical diagnosis~\cite{leibig2017leveraging,begoli2019need} and autonomous vehicles~\cite{michelmore2018evaluating}.

Another strategy is to develop AI-powered assistive technologies that can learn and adapt to user feedback over time. By allowing PVI or human assistants to correct errors and provide input to improve system performance, which is not only able to continuously improve system's reliability and robustness and also help aware users about possible AI hallucinations. This approach aligns with the principles of interactive machine learning, emphasizing the importance of human-in-the-loop learning for AI systems~\cite{amershi2014power,retzlaff2024human}.

In addition to these technical solutions, involving PVI in the design and evaluation of AI-powered assistive technologies is crucial. By engaging users as co-designers and co-evaluators, researchers and designers can gain a better understanding of the challenges and requirements of PVI, leading to the development of more reliable and robust systems. This participatory design approach has been widely advocated in the assistive technology domain~\cite{frauenberger2015pursuit,lee2004trust,zhang2023redefining}.

\section{Limitations and Conclusion}





In this study, we explored the emerging practices of Large Multimodal Model (LMM) assistance, specifically \bma, in PVI's daily lives. 
While we collected various forms of \bma's visual interpretations from participants, we allowed them to choose their preferred method of documentation.
Consequently, only 4 of the 20 collected items were screenshots that included both the pictures sent to \bma{} and its responses; the majority were textual copies, as participants found these easier to manage. 
This resulted in a limitation: our analysis lacked sufficient visual context to understand what occurred in these emerging practices, relying primarily on textual data.

Our research on \bma{} has yielded valuable insights into the potential and challenges of AI-powered assistance for PVI. By examining the experiences and interactions of PVI with \bma{}, we have identified several key themes that contribute to the ongoing discourse on the design and development of AI-powered assistive technologies. 
We explore how \bma{} facilitates task offloading through the lens of distributed cognition, enabling PVI to more effectively navigate and interact with their environment. 
Additionally, we propose design implications for AI-powered assistance based on our findings. 
By synthesizing our findings and relevant works through the lens of distributed cognition and design implications, we aim to contribute to the ongoing discourse on AI-powered assistive technologies and provide actionable insights for researchers, designers, and practitioners working in this field.

\bibliographystyle{style.bst}  
\bibliography{text}

\begin{thebibliography}{10}

\bibitem{ahmetovic2020recog}
Dragan Ahmetovic, Daisuke Sato, Uran Oh, Tatsuya Ishihara, Kris Kitani, and Chieko Asakawa.
\newblock Recog: Supporting blind people in recognizing personal objects.
\newblock In {\em Proceedings of the 2020 CHI Conference on Human Factors in Computing Systems}, pages 1--12, 2020.

\bibitem{mukhiddinov2022automatic}
Mukhriddin Mukhiddinov, Akmalbek~Bobomirzaevich Abdusalomov, and Jinsoo Cho.
\newblock Automatic fire detection and notification system based on improved yolov4 for the blind and visually impaired.
\newblock {\em Sensors}, 22(9):3307, 2022.

\bibitem{hong2022blind}
Jonggi Hong, Jaina Gandhi, Ernest~Essuah Mensah, Farnaz~Zamiri Zeraati, Ebrima Jarjue, Kyungjun Lee, and Hernisa Kacorri.
\newblock Blind users accessing their training images in teachable object recognizers.
\newblock In {\em Proceedings of the 24th International ACM SIGACCESS Conference on Computers and Accessibility}, pages 1--18, 2022.

\bibitem{morrison2023understanding}
Cecily Morrison, Martin Grayson, Rita~Faia Marques, Daniela Massiceti, Camilla Longden, Linda Wen, and Edward Cutrell.
\newblock Understanding personalized accessibility through teachable ai: Designing and evaluating find my things for people who are blind or low vision.
\newblock In {\em Proceedings of the 25th International ACM SIGACCESS Conference on Computers and Accessibility}, pages 1--12, 2023.

\bibitem{gonzalez2024investigating}
Ricardo~E Gonzalez~Penuela, Jazmin Collins, Cynthia Bennett, and Shiri Azenkot.
\newblock Investigating use cases of ai-powered scene description applications for blind and low vision people.
\newblock In {\em Proceedings of the CHI Conference on Human Factors in Computing Systems}, pages 1--21, 2024.

\bibitem{SeeingAI2020}
Microsoft.
\newblock Seeing ai - talking camera app for those with a visual impairment.
\newblock \url{https://www.microsoft.com/en-us/ai/seeing-ai}, 2022.

\bibitem{yu2023mm}
Weihao Yu, Zhengyuan Yang, Linjie Li, Jianfeng Wang, Kevin Lin, Zicheng Liu, Xinchao Wang, and Lijuan Wang.
\newblock Mm-vet: Evaluating large multimodal models for integrated capabilities.
\newblock {\em arXiv preprint arXiv:2308.02490}, 2023.

\bibitem{achiam2023gpt}
Josh Achiam, Steven Adler, Sandhini Agarwal, Lama Ahmad, Ilge Akkaya, Florencia~Leoni Aleman, Diogo Almeida, Janko Altenschmidt, Sam Altman, Shyamal Anadkat, et~al.
\newblock Gpt-4 technical report.
\newblock {\em arXiv:2303.08774}, 2023.

\bibitem{zhao2024vialm}
Yi~Zhao, Yilin Zhang, Rong Xiang, Jing Li, and Hillming Li.
\newblock Vialm: A survey and benchmark of visually impaired assistance with large models.
\newblock {\em arXiv:2402.01735}, 2024.

\bibitem{yang2024viassist}
Bufang Yang, Lixing He, Kaiwei Liu, and Zhenyu Yan.
\newblock Viassist: Adapting multi-modal large language models for users with visual impairments.
\newblock {\em arXiv:2404.02508}, 2024.

\bibitem{bma_usecase}
Announcing ``be my ai,'' soon available for hundreds of thousands of be my eyes users, 2024.

\bibitem{granquist2021evaluation}
Christina Granquist, Susan~Y Sun, Sandra~R Montezuma, Tu~M Tran, Rachel Gage, and Gordon~E Legge.
\newblock Evaluation and comparison of artificial intelligence vision aids: Orcam myeye 1 and seeing ai.
\newblock {\em Journal of Visual Impairment \& Blindness}, 115(4):277--285, 2021.

\bibitem{kupferstein2020understanding}
Elizabeth Kupferstein, Yuhang Zhao, Shiri Azenkot, and Hathaitorn Rojnirun.
\newblock Understanding the use of artificial intelligence based visual aids for people with visual impairments.
\newblock {\em Investigative Ophthalmology \& Visual Science}, 61(7):932--932, 2020.

\bibitem{BeMyEyes2020}
Be my eyes - see the world together.
\newblock \url{https://www.bemyeyes.com/}, 2024.

\bibitem{bendel2024can}
Oliver Bendel.
\newblock How can generative ai enhance the well-being of blind?
\newblock {\em arXiv:2402.07919}, 2024.

\bibitem{bigham2010vizwiz_nearly}
Jeffrey~P Bigham, Chandrika Jayant, Hanjie Ji, Greg Little, Andrew Miller, Robert~C Miller, Robin Miller, Aubrey Tatarowicz, Brandyn White, Samual White, et~al.
\newblock Vizwiz: nearly real-time answers to visual questions.
\newblock In {\em Proceedings of the 23nd annual ACM symposium on User interface software and technology}, pages 333--342, 2010.

\bibitem{bigham2010vizwiz}
Jeffrey~P Bigham, Chandrika Jayant, Andrew Miller, Brandyn White, and Tom Yeh.
\newblock Vizwiz:: Locateit-enabling blind people to locate objects in their environment.
\newblock In {\em 2010 IEEE Computer Society Conference on Computer Vision and Pattern Recognition-Workshops}, pages 65--72. IEEE, 2010.

\bibitem{burton2012crowdsourcing}
Michele~A Burton, Erin Brady, Robin Brewer, Callie Neylan, Jeffrey~P Bigham, and Amy Hurst.
\newblock Crowdsourcing subjective fashion advice using vizwiz: challenges and opportunities.
\newblock In {\em Proceedings of the 14th international ACM SIGACCESS conference on Computers and accessibility}, pages 135--142. ACM, 2012.

\bibitem{gurari2018vizwiz}
Danna Gurari, Qing Li, Abigale~J Stangl, Anhong Guo, Chi Lin, Kristen Grauman, Jiebo Luo, and Jeffrey~P Bigham.
\newblock Vizwiz grand challenge: Answering visual questions from blind people.
\newblock In {\em Proceedings of the IEEE Conference on Computer Vision and Pattern Recognition}, pages 3608--3617, 2018.

\bibitem{lasecki2013answering}
Walter~S Lasecki, Phyo Thiha, Yu~Zhong, Erin Brady, and Jeffrey~P Bigham.
\newblock Answering visual questions with conversational crowd assistants.
\newblock In {\em Proceedings of the 15th International ACM SIGACCESS Conference on Computers and Accessibility}, page~18. ACM, 2013.

\bibitem{hunaiti2006remote}
Ziad Hunaiti, Vanja Garaj, and Wamadeva Balachandran.
\newblock A remote vision guidance system for visually impaired pedestrians.
\newblock {\em The Journal of Navigation}, 59(3):497--504, 2006.

\bibitem{garaj2003system}
Vanja Garaj, Rommanee Jirawimut, Piotr Ptasinski, Franjo Cecelja, and Wamadeva Balachandran.
\newblock A system for remote sighted guidance of visually impaired pedestrians.
\newblock {\em British Journal of Visual Impairment}, 21(2):55--63, 2003.

\bibitem{baranski2015field}
Przemyslaw Baranski and Pawel Strumillo.
\newblock Field trials of a teleassistance system for the visually impaired.
\newblock In {\em 2015 8th International Conference on Human System Interaction (HSI)}, pages 173--179. IEEE, 2015.

\bibitem{bujacz2008remote}
M~Bujacz, P~Baranski, M~Moranski, P~Strumillo, and A~Materka.
\newblock Remote guidance for the blind—a proposed teleassistance system and navigation trials.
\newblock In {\em 2008 Conference on Human System Interactions}, pages 888--892. IEEE, 2008.

\bibitem{scheggi2014remote}
Stefano Scheggi, A~Talarico, and Domenico Prattichizzo.
\newblock A remote guidance system for blind and visually impaired people via vibrotactile haptic feedback.
\newblock In {\em 22nd Mediterranean Conference on Control and Automation}, pages 20--23. IEEE, 2014.

\bibitem{chaudary2017tele}
Babar Chaudary, Iikka Paajala, Eliud Keino, and Petri Pulli.
\newblock Tele-guidance based navigation system for the visually impaired and blind persons.
\newblock In {\em eHealth 360}, pages 9--16. Springer, 2017.

\bibitem{holmes2015iphone}
Nicole Holmes and Kelly Prentice.
\newblock iphone video link facetime as an orientation tool: remote o\&m for people with vision impairment.
\newblock {\em International Journal of Orientation \& Mobility}, 7(1):60--68, 2015.

\bibitem{Aira2020}
Aira.
\newblock \url{https://aira.io/}, 2024.

\bibitem{xie2022dis}
Jingyi Xie, Rui Yu, Sooyeon Lee, Yao Lyu, Syed~Masum Billah, and John~M Carroll.
\newblock Helping helpers: Supporting volunteers in remote sighted assistance with augmented reality maps.
\newblock In {\em Designing Interactive Systems Conference}, pages 881--897, 2022.

\bibitem{xie2024bubblecam}
Jingyi Xie, Rui Yu, He~Zhang, Sooyeon Lee, Syed~Masum Billah, and John~M Carroll.
\newblock Bubblecam: Engaging privacy in remote sighted assistance.
\newblock In {\em Proceedings of the CHI Conference on Human Factors in Computing Systems}, pages 1--16, 2024.

\bibitem{lasecki2011real}
Walter~S Lasecki, Kyle~I Murray, Samuel White, Robert~C Miller, and Jeffrey~P Bigham.
\newblock Real-time crowd control of existing interfaces.
\newblock In {\em Proceedings of the 24th annual ACM symposium on User interface software and technology}, pages 23--32. ACM, 2011.

\bibitem{xie2023two}
Jingyi Xie, Rui Yu, Kaiming Cui, Sooyeon Lee, John~M. Carroll, and Syed~Masum Billah.
\newblock Are two heads better than one? investigating remote sighted assistance with paired volunteers.
\newblock In {\em Proceedings of the 2023 ACM Designing Interactive Systems Conference}, DIS'23, page 1810–1825, 2023.

\bibitem{kamikubo2020support}
Rie Kamikubo, Naoya Kato, Keita Higuchi, Ryo Yonetani, and Yoichi Sato.
\newblock Support strategies for remote guides in assisting people with visual impairments for effective indoor navigation.
\newblock In {\em Proceedings of the 2020 CHI Conference on Human Factors in Computing Systems}, pages 1--12, 2020.

\bibitem{c4vtochi}
Jingyi Xie, Madison Reddie, Sooyeon Lee, Syed~Masum Billah, Zihan Zhou, Chun-hua Tsai, and John~M Carroll.
\newblock Iterative design and prototyping of computer vision mediated remote sighted assistance.
\newblock {\em ACM Transactions on Computer-Human Interaction (TOCHI)}, 29(4):1--40, 2022.

\bibitem{iui}
Sooyeon Lee, Rui Yu, Jingyi Xie, Syed~Masum Billah, and John~M Carroll.
\newblock Opportunities for human-ai collaboration in remote sighted assistance.
\newblock In {\em 27th International Conference on Intelligent User Interfaces}, pages 63--78, 2022.

\bibitem{lee2020emerging}
Sooyeon Lee, Madison Reddie, Chun-Hua Tsai, Jordan Beck, Mary~Beth Rosson, and John~M Carroll.
\newblock The emerging professional practice of remote sighted assistance for people with visual impairments.
\newblock In {\em Proceedings of the 2020 CHI Conference on Human Factors in Computing Systems}, pages 1--12, 2020.

\bibitem{Caroll2020Human}
John~M. Carroll, Sooyeon Lee, Madison Reddie, Jordan Beck, and Mary~Beth Rosson.
\newblock Human-computer synergies in prosthetic interactions.
\newblock {\em IxD{\&}A}, 44:29--52, 2020.

\bibitem{lee2018conversations}
Sooyeon Lee, Madison Reddie, Krish Gurdasani, Xiying Wang, Jordan Beck, Mary~Beth Rosson, and John~M. Carroll.
\newblock Conversations for vision: Remote sighted assistants helping people with visual impairments, 2018.

\bibitem{hutchins1995cognition}
Edwin Hutchins.
\newblock {\em Cognition in the Wild}.
\newblock MIT press, 1995.

\bibitem{hollan2000distributed}
James Hollan, Edwin Hutchins, and David Kirsh.
\newblock Distributed cognition: toward a new foundation for human-computer interaction research.
\newblock {\em ACM Transactions on Computer-Human Interaction (TOCHI)}, 7(2):174--196, 2000.

\bibitem{rogers2022hci}
Yvonne Rogers.
\newblock {\em HCI theory: classical, modern, and contemporary}.
\newblock Springer Nature, 2022.

\bibitem{halverson2002activity}
Christine~A Halverson.
\newblock Activity theory and distributed cognition: Or what does cscw need to do with theories?
\newblock {\em Computer Supported Cooperative Work (CSCW)}, 11:243--267, 2002.

\bibitem{baumer2011comparing}
Eric~PS Baumer and Bill Tomlinson.
\newblock Comparing activity theory with distributed cognition for video analysis: beyond" kicking the tires".
\newblock In {\em Proceedings of the SIGCHI conference on human factors in computing systems}, pages 133--142, 2011.

\bibitem{wu2008collaborating}
Mike Wu, Jeremy Birnholtz, Brian Richards, Ronald Baecker, and Mike Massimi.
\newblock Collaborating to remember: a distributed cognition account of families coping with memory impairments.
\newblock In {\em Proceedings of the SIGCHI conference on human factors in computing systems}, pages 825--834, 2008.

\bibitem{rogers2006distributed}
Yvonne Rogers.
\newblock Distributed cognition and communication.
\newblock 2006.

\bibitem{perry2003distributed}
Mark Perry.
\newblock Distributed cognition.
\newblock {\em HCI models, theories, and frameworks: Toward a multidisciplinary science}, pages 193--223, 2003.

\bibitem{nobarany2012facilitating}
Syavash Nobarany, Mona Haraty, and Brian Fisher.
\newblock Facilitating the reuse process in distributed collaboration: a distributed cognition approach.
\newblock In {\em Proceedings of the ACM 2012 conference on Computer Supported Cooperative Work}, pages 1223--1232, 2012.

\bibitem{braun2006using}
Virginia Braun and Victoria Clarke.
\newblock Using thematic analysis in psychology.
\newblock {\em Qualitative research in psychology}, 3(2):77--101, 2006.

\bibitem{avila2016remote}
Mauro Avila, Katrin Wolf, Anke Brock, and Niels Henze.
\newblock Remote assistance for blind users in daily life: A survey about be my eyes.
\newblock In {\em Proceedings of the 9th ACM International Conference on PErvasive Technologies Related to Assistive Environments}, pages 1--2, 2016.

\bibitem{BME_100ways}
100 ways to use be my eyes.
\newblock \url{https://support.bemyeyes.com/hc/en-us/articles/360005528317-100-Ways-to-Use-Be-My-Eyes}, 2024.

\bibitem{truhn2023large}
Daniel Truhn, Jorge~S Reis-Filho, and Jakob~Nikolas Kather.
\newblock Large language models should be used as scientific reasoning engines, not knowledge databases.
\newblock {\em Nature medicine}, 29(12):2983--2984, 2023.

\bibitem{zhu2014reasoning}
Yuke Zhu, Alireza Fathi, and Li~Fei-Fei.
\newblock Reasoning about object affordances in a knowledge base representation.
\newblock In {\em Computer Vision--ECCV 2014: 13th European Conference, Zurich, Switzerland, September 6-12, 2014, Proceedings, Part II 13}, pages 408--424. Springer, 2014.

\bibitem{chen2008using}
Liming Chen, Chris Nugent, Maurice Mulvenna, Dewar Finlay, Xin Hong, and Michael Poland.
\newblock Using event calculus for behaviour reasoning and assistance in a smart home.
\newblock In {\em Smart Homes and Health Telematics: 6th International Conference, ICOST 2008 Ames, IA, USA, June 28-July 2, 2008 Proceedings 6}, pages 81--89. Springer, 2008.

\bibitem{fischer2001user}
Gerhard Fischer.
\newblock User modeling in human--computer interaction.
\newblock {\em User modeling and user-adapted interaction}, 11:65--86, 2001.

\bibitem{amershi2014power}
Saleema Amershi, Maya Cakmak, William~Bradley Knox, and Todd Kulesza.
\newblock Power to the people: The role of humans in interactive machine learning.
\newblock {\em AI magazine}, 35(4):105--120, 2014, doi:https://doi.org/10.1609/aimag.v35i4.2513.

\bibitem{huffman1993goal}
Cynthia Huffman and Michael~J Houston.
\newblock Goal-oriented experiences and the development of knowledge.
\newblock {\em Journal of Consumer Research}, 20(2):190--207, 1993.

\bibitem{letier2002agent}
Emmanuel Letier and Axel Van~Lamsweerde.
\newblock Agent-based tactics for goal-oriented requirements elaboration.
\newblock In {\em Proceedings of the 24th international conference on Software Engineering}, pages 83--93, 2002.

\bibitem{liu2008camera}
Xu~Liu.
\newblock A camera phone based currency reader for the visually impaired.
\newblock In {\em Proceedings of the 10th international ACM SIGACCESS conference on Computers and accessibility}, pages 305--306, 2008.

\bibitem{parlouar2009assistive}
R{\'e}mi Parlouar, Florian Dramas, Marc~MJ Mac{\'e}, and Christophe Jouffrais.
\newblock Assistive device for the blind based on object recognition: an application to identify currency bills.
\newblock In {\em Proceedings of the 11th international ACM SIGACCESS conference on Computers and accessibility}, pages 227--228, 2009.

\bibitem{paisios2012exchanging}
Nektarios Paisios, Alexander Rubinsteyn, and Lakshminarayanan Subramanian.
\newblock Exchanging cash with no fear: A fast mobile money reader for the blind.
\newblock In {\em Workshop on Frontiers in Accessibility for Pervasive Computing. ACM}, 2012.

\bibitem{girshick2014rich}
Ross Girshick, Jeff Donahue, Trevor Darrell, and Jitendra Malik.
\newblock Rich feature hierarchies for accurate object detection and semantic segmentation.
\newblock In {\em Proceedings of the {IEEE} {C}onference on {C}omputer {V}ision and {P}attern {R}ecognition}, pages 580--587, 2014.

\bibitem{girshick2015fast}
Ross Girshick.
\newblock Fast r-cnn.
\newblock In {\em Proceedings of the {IEEE} {I}nternational {C}onference on {C}omputer {V}ision}, pages 1440--1448, 2015.

\bibitem{ren2016faster}
Shaoqing Ren, Kaiming He, Ross Girshick, and Jian Sun.
\newblock Faster r-cnn: Towards real-time object detection with region proposal networks.
\newblock {\em IEEE transactions on pattern analysis and machine intelligence}, 39(6):1137--1149, 2016.

\bibitem{krizhevsky2017imagenet}
Alex Krizhevsky, Ilya Sutskever, and Geoffrey~E Hinton.
\newblock Imagenet classification with deep convolutional neural networks.
\newblock {\em Communications of the ACM}, 60(6):84--90, 2017.

\bibitem{ma2018arbitrary}
Jianqi Ma, Weiyuan Shao, Hao Ye, Li~Wang, Hong Wang, Yingbin Zheng, and Xiangyang Xue.
\newblock Arbitrary-oriented scene text detection via rotation proposals.
\newblock {\em IEEE Transactions on Multimedia}, 20(11):3111--3122, 2018.

\bibitem{he2017deep}
Wenhao He, Xu-Yao Zhang, Fei Yin, and Cheng-Lin Liu.
\newblock Deep direct regression for multi-oriented scene text detection.
\newblock In {\em Proceedings of the IEEE International Conference on Computer Vision}, pages 745--753, 2017.

\bibitem{zhou2017east}
Xinyu Zhou, Cong Yao, He~Wen, Yuzhi Wang, Shuchang Zhou, Weiran He, and Jiajun Liang.
\newblock East: an efficient and accurate scene text detector.
\newblock In {\em Proceedings of the IEEE conference on Computer Vision and Pattern Recognition}, pages 5551--5560, 2017.

\bibitem{yao2016scene}
Cong Yao, Xiang Bai, Nong Sang, Xinyu Zhou, Shuchang Zhou, and Zhimin Cao.
\newblock Scene text detection via holistic, multi-channel prediction.
\newblock {\em arXiv preprint arXiv:1606.09002}, 2016.

\bibitem{liu2017deep}
Yuliang Liu and Lianwen Jin.
\newblock Deep matching prior network: Toward tighter multi-oriented text detection.
\newblock In {\em Proceedings of the IEEE Conference on Computer Vision and Pattern Recognition}, pages 1962--1969, 2017.

\bibitem{lyu2018multi}
Pengyuan Lyu, Cong Yao, Wenhao Wu, Shuicheng Yan, and Xiang Bai.
\newblock Multi-oriented scene text detection via corner localization and region segmentation.
\newblock In {\em Proceedings of the IEEE conference on computer vision and pattern recognition}, pages 7553--7563, 2018.

\bibitem{leibig2017leveraging}
Christian Leibig, Vaneeda Allken, Murat~Se{\c{c}}kin Ayhan, Philipp Berens, and Siegfried Wahl.
\newblock Leveraging uncertainty information from deep neural networks for disease detection.
\newblock {\em Scientific reports}, 7(1):1--14, 2017, doi:https://doi.org/10.1038/s41598-017-17876-z.

\bibitem{begoli2019need}
Edmon Begoli, Tanmoy Bhattacharya, and Dimitri Kusnezov.
\newblock The need for uncertainty quantification in machine-assisted medical decision making.
\newblock {\em Nature Machine Intelligence}, 1(1):20--23, 2019, doi:https://doi.org/10.1038/s42256-018-0004-1.

\bibitem{michelmore2018evaluating}
Rhiannon Michelmore, Marta Kwiatkowska, and Yarin Gal.
\newblock Evaluating uncertainty quantification in end-to-end autonomous driving control, 2018.

\bibitem{retzlaff2024human}
Carl~Orge Retzlaff, Srijita Das, Christabel Wayllace, Payam Mousavi, Mohammad Afshari, Tianpei Yang, Anna Saranti, Alessa Angerschmid, Matthew~E Taylor, and Andreas Holzinger.
\newblock Human-in-the-loop reinforcement learning: A survey and position on requirements, challenges, and opportunities.
\newblock {\em Journal of Artificial Intelligence Research}, 79:359--415, 2024.

\bibitem{frauenberger2015pursuit}
Christopher Frauenberger, Judith Good, Geraldine Fitzpatrick, and Ole~Sejer Iversen.
\newblock In pursuit of rigour and accountability in participatory design.
\newblock {\em International journal of human-computer studies}, 74:93--106, 2015.

\bibitem{lee2004trust}
John~D Lee and Katrina~A See.
\newblock Trust in automation: Designing for appropriate reliance.
\newblock {\em Human factors}, 46(1):50--80, 2004.

\bibitem{zhang2023redefining}
He~Zhang, Chuhao Wu, Jingyi Xie, Yao Lyu, Jie Cai, and John~M. Carroll.
\newblock Redefining qualitative analysis in the ai era: Utilizing chatgpt for efficient thematic analysis, 2023.

\end{thebibliography}

\end{document}